\documentclass[iop]{emulateapj}

\usepackage{natbib}

\shorttitle{Deformation of Optics}
\shortauthors{Peterson et al.}

\begin{document}

\title{Deformation of Optics for Photon Monte Carlo Simulations}

\author{J.~R.~Peterson$^1$, E.~Peng$^1$, C.~J.~Burke$^1$, G.~Sembroski$^1$, J.~Cheng$^1$,}

\affil{$^1$ Department of Physics and Astronomy, Purdue University, West Lafayette, IN 47907}

\email{peters11@purdue.edu}

\begin{abstract}

We develop a comprehensive approach to simulate the deformation of mirrors and lenses due to thermal and mechanical stresses that couples efficiently to photon-based optics simulations.  This expands upon previous work where we demonstrated a comprehensive ab initio approach to simulate astronomical images using a photon Monte Carlo method.  We apply elasticity theory and estimate thermal effects by adapting a three-dimensional numerical method.  We also consider the effect of active optics control systems and active cooling systems in further correcting distortions in the optics.  We validate the approach by showing convergence to analytic estimates, and then apply the methodology to the WIYN 3.5m telescope primary mirror.  We demonstrate that changes in the soak temperature result in second order point spread function (PSF) defocusing, the gravitational sag and positioning errors result in highly structured PSF distortions, and large-scale thermal gradients result in an elliptical PSF distortion patterns.  All three aspects of the environment are larger than the intrinsic optical aberrations of the design, and further exploration with a variety of telescopes should lead to detailed PSF size and shape, astrometric distortion, and field variation predictions.   The simulation capabilities developed in this work is publicly available with the Photon Simulation (PhoSim) package.

\end{abstract}


\section{Introduction}

Simulations have become an indispensable part of modern science.  In astrophysics, simulations are used extensively to simulate astronomical objects (e.g. \citealt{fryxell2000}; \citealt{springel2005}; \citealt{stone2008}) and simulations of observations are critical in interpreting observations in high energy (e.g \citealt{peterson2004}; \citealt{peterson2007}; \citealt{andersson2007}; \citealt{davis2012}; \citealt{ackermann2012}) and optical astrophysics (e.g. \citealt{lane}, \citealt{ellerbroek}; \citealt{lelouarn}; \citealt{britton}; \citealt{jolissaint}; \citealt{bertin}; \citealt{dobke}; \citealt{peterson2015}) .  In a previous work (\citealt{peterson2015}), we developed a comprehensive ab initio simulation approach using a photon Monte Carlo methodology in order to simulate high fidelity astronomical images from optical and infrared survey telescopes.  In \citealt{peterson2015}, we noted that PhoSim is capable of including arbitrary effects of this nature to affect the path of the light as the geometric optics part of the simulation was done.  However, we did not specify how to generate realistic shapes that are consistent with the physics of the deformation of optics due to mechanical and thermal stresses.
Hence, it is important to study this in further detail.  In addition, these kinds of perturbations on the optical design are not just a small complication.  With modern large telescopes mirrors, optics shape distortion can be the single most important effect on the PSF structure and pattern (\citealt{peterson2015}).

Although the deformation of optics has been extensively studied by telescope designers and builders (e.g. \citealt{angel1984}), very little is known about the direct consequence on astronomical results.  The combination of the physics simulation in this work and the generalized photon simulation framework in PhoSim can address these more systematically.  Generally, the optics are a significant component of the PSF.  The design itself can be a non-negligible component of the PSF and is often studied extensively in the construction of a new telescope, but the perturbations due to thermal and mechanical effects often exceed the design by a significant amount (\citealt{peterson2015}).  For ground-based telescopes the goal is often to keep the instrumental PSF below that of the typical atmospheric seeing.  For wide-field telescopes with active optics system it is often slightly below the atmospheric PSF, since the atmosphere itself can be the primary noise source in sensing the optical surface distortions.  For space-based instruments, the optics PSF can dominate.
Similarly, the ellipticity of the PSF can be dominated by the distortion of optics even though the PSF may be smaller.  This is because the optical deformation PSF is often highly elliptical compared to the atmospheric PSF due to the nature of the deformations (\citealt{chang2013}).  In addition, the astrometric distortion on large angular scales (several arcminutes) is also due to the optical distortions.  However, all of these qualitative conclusions are poorly understood and deserves more quantitative investigation.

We are particularly interested in making a fundamental numerical approach that considers the opto-mechanical physics in a coherent fashion that will naturally work as a part of complete astronomical image simulations.  As indicated above, our previous work accepted arbitrary surface perturbations to place on individual optics.  In turn, those patterns could be predicted from comprehensive finite element engineering codes of the optics.  However, that method is only useful an isolated individual simulations and not useful generally for a suite of comprehensive astronomical pipeline studies or for investigating more dynamic effects (i.e. control systems).  Consequently, we develop a comprehensive numerical methodology that will naturally couple efficiently to a photon Monte Carlo.  This work is organized as follows.  In \S1, we describe this numerical approach.  In \S2, we estimate the magnitude of the thermal environmental effects.  In \S4, we validate these results with analytic predictions for a thin plate.  In \S5, we consider an active optics control system.   In \S6, we describe the opto-mechanical coupling.  In \S7, we apply this method to the primary mirror of the WIYN 3.5m telescope and discuss our conclusions in \S8.

\section{Deformation of Optics}

In order to accurately simulate the shapes of optics, we calculate deformations via elasticity theory developed by Cauchy and Poisson (see e.g. \cite{landau1986}).   We constructed a code to use the distinct lattice spring method (DLSM) developed by \citealt{zhao2011}.  This is similar to finite element methods, but represents a solid as a series of particles in a lattice connected by springs.  This method is both computationally efficient, and naturally well-coupled to photon raytracing because it also represents a surface by a series of points that we will use with the photon simulations later.

We then represent optics in the following manner.  We construct a three-dimensional cubic lattice with $N$ points on each side.  An optic is assumed to be cylindrically-symmetric and is represented by considering the points that would be contained interior to the 3-dimensional volume.  The three-dimensional volume is specified entirely by the two aspherical surface in the case of a lens.  Conversely, in the case of a mirror additional information about the thickness and the shape of the back surface is an additional input.  We also allow for fused optics and then the top surface is specified by both aspheric surfaces and a boundary area.  We then make the three dimensional lattice physical point separation, $d$ equal to $\frac{2L}{N-1}$ where $L$ is the outer radius of the optics.  We then loop through all points to specify which points are contained within the theoretical three-dimensional volume.  In general, choosing a large value of $N$ results in greater accuracy, but we have found any value of $N$ will produce reasonable results unless $N$ is so small that the number of points representing the thickness of the optic is less than 3.

Generally, we are interested in the exact details of the shape of either top or bottom surfaces.  We found that numerically the best way to approach this is to then perturb the points near the surface from their nominal lattice position to match the exact surface exactly.  Otherwise, the final perturbed surface may have a blocky numerical residual structure for finite values of $N$.  This representation become more accurate at larger values of $N$.  The DLSM method is then slightly modified by the different pairs lattice points having a larger nominal separation than their nominal location.

For every point, we calculate the number of connecting points in the lattice.  For points in the middle of the optic, this is always 18 points, since \citealt{zhao2011} found that the most robust results would come from considering all the adjacent points (6) and the diagonal points in the x, y, and z planes (12).  For points near the edge and sides it will be less than 18.  We further compute the value of $\alpha_N$ locally which is used to scale the elastic forces as in \citealt{zhao2011}.

For every optic, we consider how it is held in place.  We allow for three types of support:  1) the optic is supported by using actuators or supports on the bottom, 2) the optic is supported by actuators or supports on the top, and 3) supported by points on the side.  For mounting from the top and bottom, we use a series of actuators or supports placed in a pattern.  Currently we have implemented three patterns with constant spacing:  1) a staggered grid pattern, 2) an outside rim pattern, and 3) a concentric ring pattern.  Other patterns are straight-forward to implement.
For each actuator, the radial footprint describes the total region where there is an applied force.  Similarly, for the side mounting we consider locations at the edge of the cylinder at the midpoint of the vertical.  We then also have a circular footprint for each support points.  Thus, we can either have the optic supported by a continuous radial support or it could be supported by a small number of distinct azimuthal positions.  Then, for all support schemes, we flag the subset of all possible points representing the optic where they would be contained within a support actuator or constraint.  These points will either be considered immobile in the physics calculation below or will be moved by the telescope control system to a distinct new location.  Therefore, the other points in the optic will be forced to respond to these constraints and pulled in the appropriate direction through the internal forces.

We then calculate the local temperature, $T_l$ of a point ($x$, $y$, $z$) by

$$ T_l = T_s + x \frac{dT}{dr} sin(\theta_g) cos(\phi_g) + y \frac{dT}{dr} sin(\theta_g) sin(\phi_g)$$
$$+ z \frac{dT}{dr} cos(\theta_g) - T_0 $$

\noindent
where $T_s$ is the sink temperature of the environment, $\frac{dT}{dr}$ is the radial temperature gradient, $\theta_g$ is the polar angle of the temperature gradient, $\phi_g$ is the azimuthal angle of the temperature gradient, and $T_0$ is the temperature of the optic when the surface is figured.  To simulate the thermal expansion, all the lattice points are then dilated from their nominal position by $\vec{r}^{\prime} = \vec{r} \left( 1 + \alpha T_l \right)$.  This new position is then considered the equilibrium position for the elastic simulation below.  This is a reasonable methodology since the time scale for thermal conduction (hours) is much longer than the adjustment to elastic or gravitational forces (tiny fractions of a second).  Note the coordinates are carefully chosen to make the origin equal to the center of mass of the optic.

For the elasticity calculation, we consider the distortion of the original position in the lattice ($x$, $y$, $z$) to a possible new position, ($x^{\prime}$, $y^{\prime}$, $z^{\prime}$).  The goal of the calculation is to iteratively adjust the position until the optic is in equilibrium.  We start by calculating the strain tensor for each point according to

$$ \epsilon_{ij} = \frac{1}{2} \left( \frac{du_i}{dx_j} + \frac{du_j}{dx_j} \right) $$

\noindent
where $x_1, x_2, x_3$ = $x, y, z$ and $u_1, u_2, u_3$ = $x_1^{\prime} - x_1, x_2^{\prime} - x_2, x_3^{\prime} - x_3$.  The force on a particular node point due to the elastic deformation of each of its neighbors is then given by

$$ F_i = k_n \vec{u}_{12} \cdot \hat{n} \hat{n_i} + k_s ( \bar{e} \hat{n_x} - \bar{e} \cdot \hat{n} \hat{n_i} )$$

\noindent
where $\vec{u}_{12} = \vec{u}_1 - \vec{u}_2$ and

$$\bar{e} = \frac{1}{2} (\epsilon_{1ij} + \epsilon_{2ij} ) \cdot \hat{n} \cdot \vline \vec{x}_1 - \vec{x}_2 \vline$$

\noindent
The elastic force coefficients are given by

$$ k_n = \frac{3Ed}{(1 - 2 \nu) \alpha_N} $$

$$ k_s = \frac{3 ( 1 - 4 \nu ) Ed}{(1+\nu) ( 1 - 2 \nu) \alpha_N}$$

\noindent
where $E$ is the Young's modulus and $\nu$ is the Poisson ratio of the optical material.  The forces of all the adjacent node points (up to 18) are summed.  Gravity is then included by adding a force

$$ F_x = - m g sin(\theta) cos(\phi)$$
$$ F_y = - m g sin(\theta) sin(\phi)$$
$$ F_z = - m g cos(\theta)$$

Most likely the optic was figured straight up, so then we add a counter force equal to $ F_z = m g $, so that there is no net gravitational force when the optic is placed in the same configuration.  If it was figured in a different way or it was figured on Earth and put into orbit then the corresponding force can be configured differently.  We then update the velocity of each point by taking the net force divide by the mass and multiply by the time step which we choose to be

$$ \delta t = f_1 \sqrt{\frac{E d}{3 ( 1 - 2 \nu)} + \frac{4}{3} \frac{E d}{2 ( 1 + \nu)}}$$

\noindent
which is the inverse of the natural oscillation frequency.  We choose $f_1$ to be 0.1 which produces stable numerical results.  We also attenuate the velocity by $f_2 \mid F_i \mid \frac{dt}{m} \frac{v_i}{\mid v_i \mid}$ to settle the points into equilibrium by adding artificial friction.  Choosing $f_2$ to be 0.8 produces robust results as in \citealt{zhao2011}.

During the calculation, we may modify the positions of any movable actuators as a results of the control system attempting to modify the surface shape.  Similarly, the constrained points are not allowed to move in the simulation.  We can relate these constraints to an equivalent applied force.  Finally, the elasticity calculation ends when the average velocity of the lattice points times the time step is less than the desired surface step tolerance.

\section{Coupling to Environment}

The three most important environmental consideration that will affect the optics are:  1) the gravitational field, 2) the overall soak temperature of the system, and 3) the presence of temperature non-uniformities and gradients.  This first is straight-forward, since the angle from zenith is the only relevant variable and is implemented by the gravitational force calculation above.

To estimate the soak temperature for a given time, we consider a class of diurnal empirical models.  Although the typical temperature pattern will be driven by the complexity of weather and by local thermal variations, the average temperature behavior is well-modelled by parameterization like \citealt{parton1981}. \citealt{duan2012} reviewed six variants on the \citealt{parton1981} comparing with measurements and the simplest was \citealt{schlaedich2001}.  Here the temperature is given by

$$  T = T_l + (T_h - T_l) \cos{\frac{\pi}{\omega} (t_s - t_h)} e^{-\frac{t - t_s}{k}}$$

$$ T = T_l + (T_h - T_l) \cos{\pi/\omega (t - t_h)}$$

$$ k = \frac{\omega}{\pi} \arctan{\frac{\pi}{\omega} (t_s - t_h)} $$

\noindent
where $T_l$ is the low temperature, $T_h$ is the daily high temperature, $\omega$ is the hours of daylight, $T_s$ is the time of sunset.  There is a significant lag for large optics to come into thermal equilibrium, so we can reference the appropriate lagged temperature.

Temperature gradients and non-uniformities depend on both the timescale that the optic reaches thermal equilibrium as well as the thermal history of the environment around the optic.  This has been considered previously to estimate the opto-mechanical performance and thermal environment of some systems (\citealt{banyal2013}, \citealt{gracey2016}, \citealt{ek2018}).  Since we are particularly interested in the thermal gradients and asymmetry patterns, this ultimately depends on the uneven heating and cooling of the surrounding structures and airflow patterns.  Rather than model this full external environment which would be beyond the scope of this work and may not even be fully understood at a given site, we can estimate the typical magnitude of thermal gradients based on the thermal history of the environment.

Since most optics are relatively thin in one-dimension consider that an infinite plate has a cooling time-scale given by:

$$ t_0 = \frac{h^2 \rho C_P}{\pi^2 \kappa} $$

\noindent
where $h$ is the thickness, $\rho$ is the density, $C_p$ is the heat capacity, and $\kappa$ is the thermal conductivity (e.g. \citealt{carslaw1959}).  For the thickness, $h$, we use the volume averaged thickness, and the others are given by the material properties of the mirror.  This then estimates the e-folding time-scale for an optic to respond to its environment.   The change in temperature, $\Delta T$, during the equilibration is then given by $\Delta T = \frac{dT}{dt} t_0$.  Then the temperature gradient, $\frac{dT}{dx}$ would be given by

$$ \frac{dT}{dx}  = \frac{dT}{dt} \frac{t_0}{L} $$

\noindent
where $L$ is the outer radius of the optic.  Thus, although the exact value of the gradient is difficult to predict without complete thermal modelling of the whole environment, typical thermal gradient should be proportional to all these factors.  Note that the linear dependence on the environment's thermal derivative, $\frac{dT}{dt}$, implies that thermal gradients are more significant when the temperature is changing rapidly (shortly after sunset).  This thermal derivative is then calculable from the temperature evolution above.

Some large mirror systems have cooling systems to try to minimize temperature gradients.  A straight-forward way of approximating this is to describe this as an increase in thermal conductivity, since it will result in a larger heat transfer than without it.  Therefore, an effective conductivity due to the cooling or heating system can be estimated as

$$ \kappa_c = \frac{d\dot{Q}}{dT} \frac{h}{\pi L^2} $$

\noindent
where $\frac{d\dot{Q}}{dT}$ is the power per temperature of the conditioning system, $h$ is the effective height of the optic, and $L$ is the outer radius.  This can modify the expression above to increase the conductivity above passive heat transfer.

\section{Basic Results and Validation}

\begin{figure*}[htb]
\begin{center}
\includegraphics[width=0.99\columnwidth]{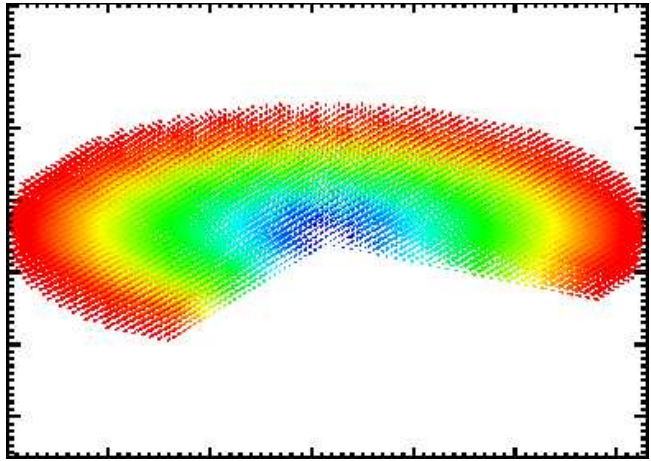}
\end{center}
\caption{\label{fig:label1}  Three-dimensional vector plot of the displacements of the points representing the glass cylindrical optic.  The colors are proportional to the magnitude of the displacement (purple=smallest and red=largest).  The optic is supported from the outer rim.  The lattice points are responding to an external temperature 10 degrees below the temperature the optic was constructed.  The vectors are indicating the optic is expanding radially and vertically (less visible).  The numerical value of the expansion is compared in Figures~\ref{fig:label2} and~\ref{fig:label3}.  }
\end{figure*}

\begin{figure*}[htb]
\begin{center}
\includegraphics[width=0.99\columnwidth]{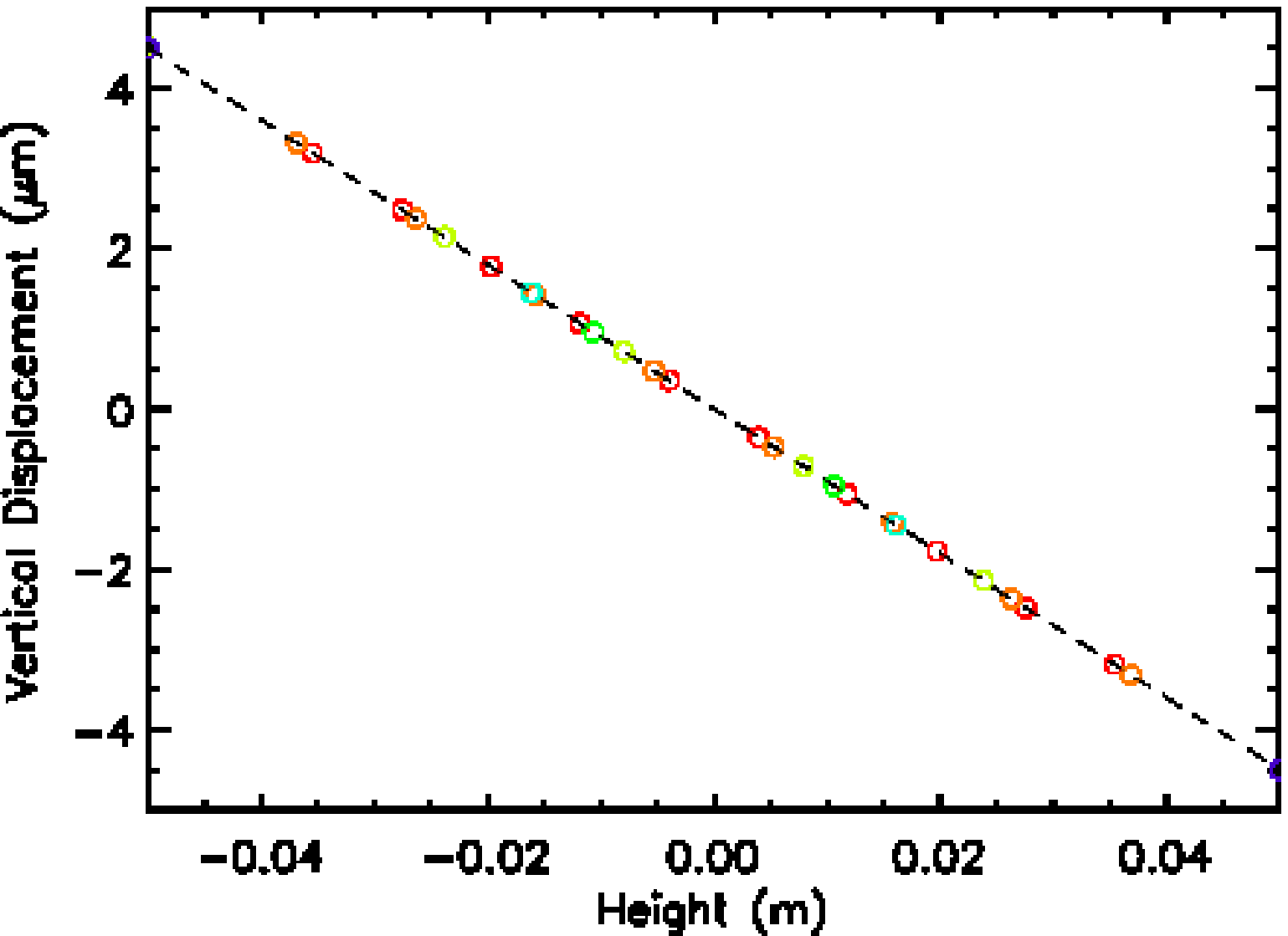}
\end{center}
\caption{\label{fig:label2}  The vertical displacement as a function of height in the optic show in Figure~\ref{fig:label2} and described in the text with a uniform soak temperature.  The colors (purple to red) represent the code run with various numerical resolutions (16,24,32,48,64,96,128).  The points agree with the analytic solution of linear expansion (dashed line) and any error is less than 10 $nm$ for $N\geq16$.}
\end{figure*}

\begin{figure*}[htb]
\begin{center}
\includegraphics[width=0.99\columnwidth]{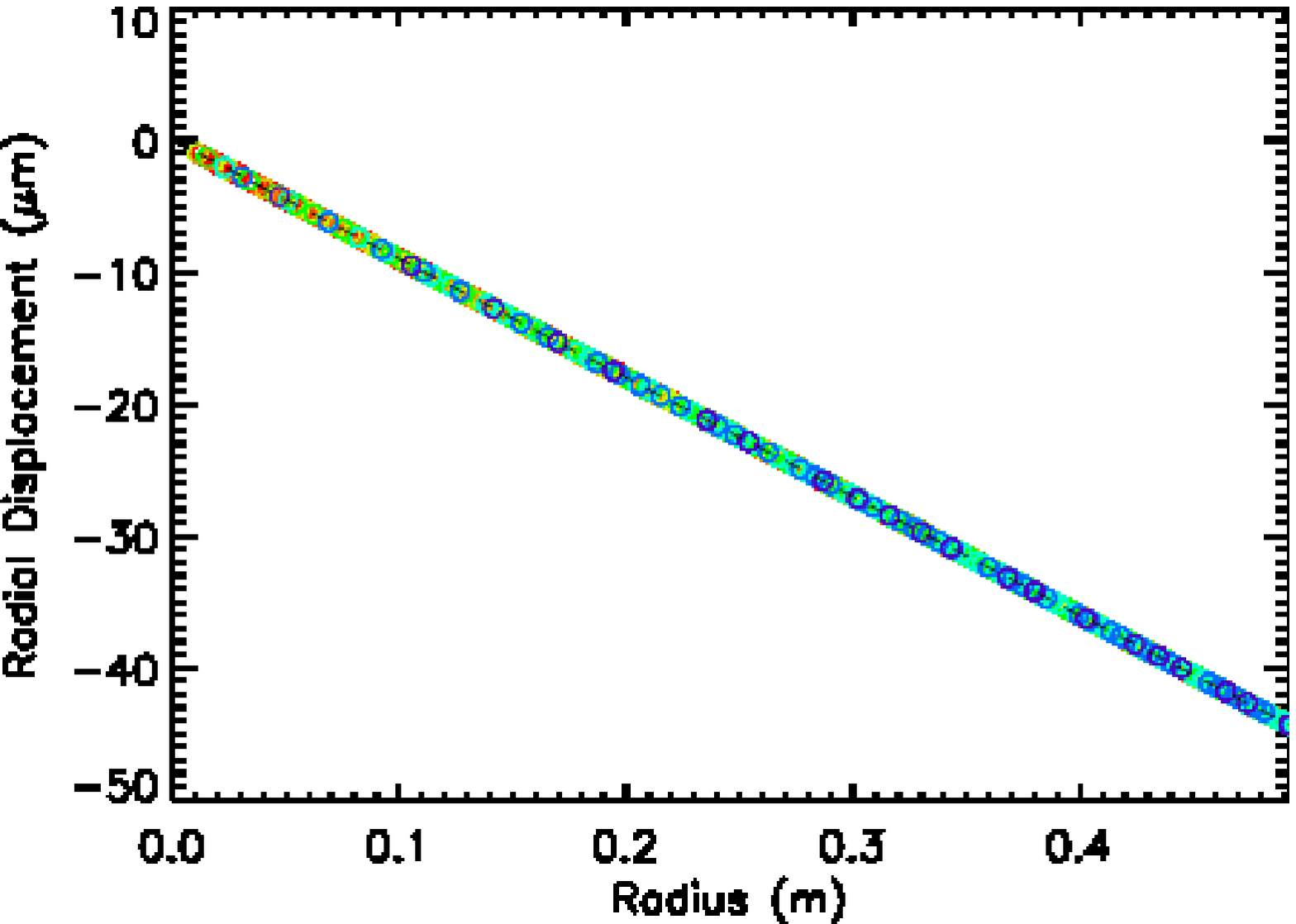}
\end{center}
\caption{\label{fig:label3}  The radial displacement as a function of radius in the optic shown in Figure~\ref{fig:label2} and described in the text with a uniform soak temperature.  The colors (purple to red) represent the code run with various numerical resolutions (16,24,32,48,64,96,128).  The points agree with the analytic solution of linear expansion (dashed line) and any error is less than 10 $nm$ for $N\geq16$.}
\end{figure*}

\begin{figure*}[htb]
\begin{center}
\includegraphics[width=0.99\columnwidth]{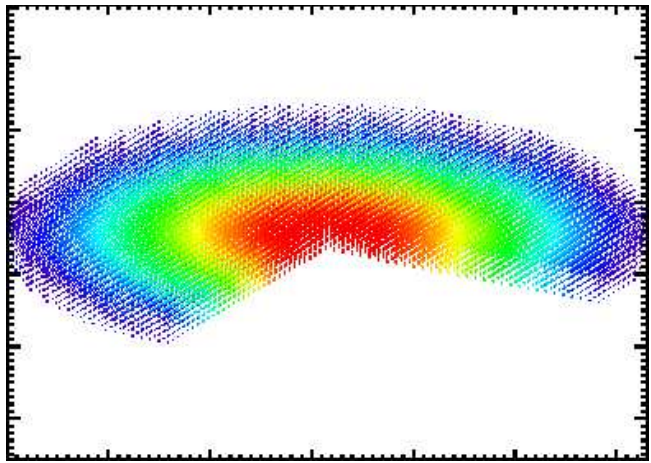}
\end{center}
\caption{\label{fig:label4}  Three-dimensional vector plot of the displacements of the points representing the glass cylindrical optic.  The colors are proportional to the magnitude of the displacement (purple=smallest and red=largest).  The optic is supported from the outer rim.  In this figure, the points are responding to a uniform gravitational field in the vertical direction, and the center regions have the largest sag.  The numerical value of the sag as a function of radius is shown in~\ref{fig:label5}.}
\end{figure*}

\begin{figure*}[htb]
\begin{center}
\includegraphics[width=0.99\columnwidth]{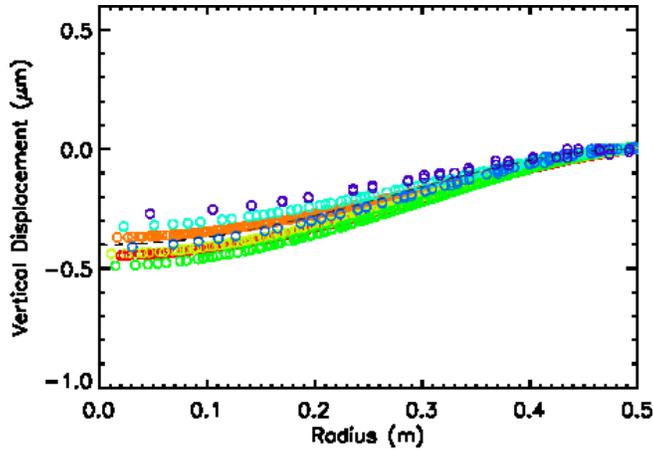}
\end{center}
\caption{\label{fig:label5}  The amount of sag as a function of radius of the optic shown in Figure~\ref{fig:label1} and described in the text due to gravity.  The colors (purple to red) represent the code run with various numerical resolutions (16,24,32,48,64,96,128).  The points asymptotically approach the analytic solution of Poisson for a circular plate supported at the edges (dashed line).  The error is less than 25 $nm$ for $N\geq 64$.}
\end{figure*}

\begin{figure*}[htb]
\begin{center}
\includegraphics[width=0.99\columnwidth]{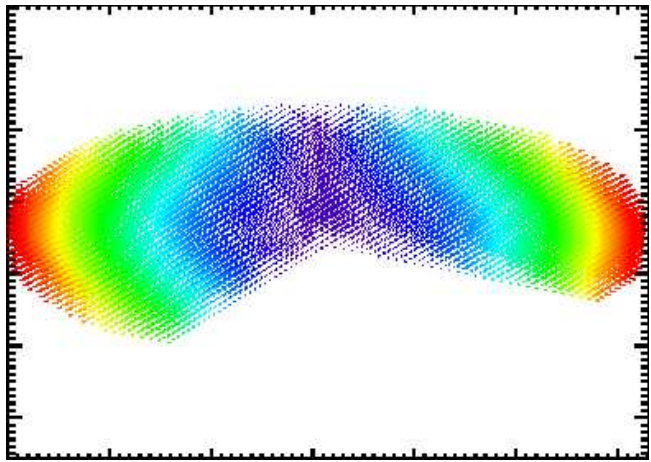}
\end{center}
\caption{\label{fig:label6}  Three-dimensional vector plot of the displacements of the points representing the glass cylindrical optic.  The colors are proportional to the magnitude of the displacement (purple=smallest and red=largest).  The optic is supported from the outer rim.  The lattice points are responding to a temperature gradient due to a change of temperature derivative of $0.4^{\circ}$C per hour.}
\end{figure*}

\begin{figure*}[htb]
\begin{center}
\includegraphics[width=0.99\columnwidth]{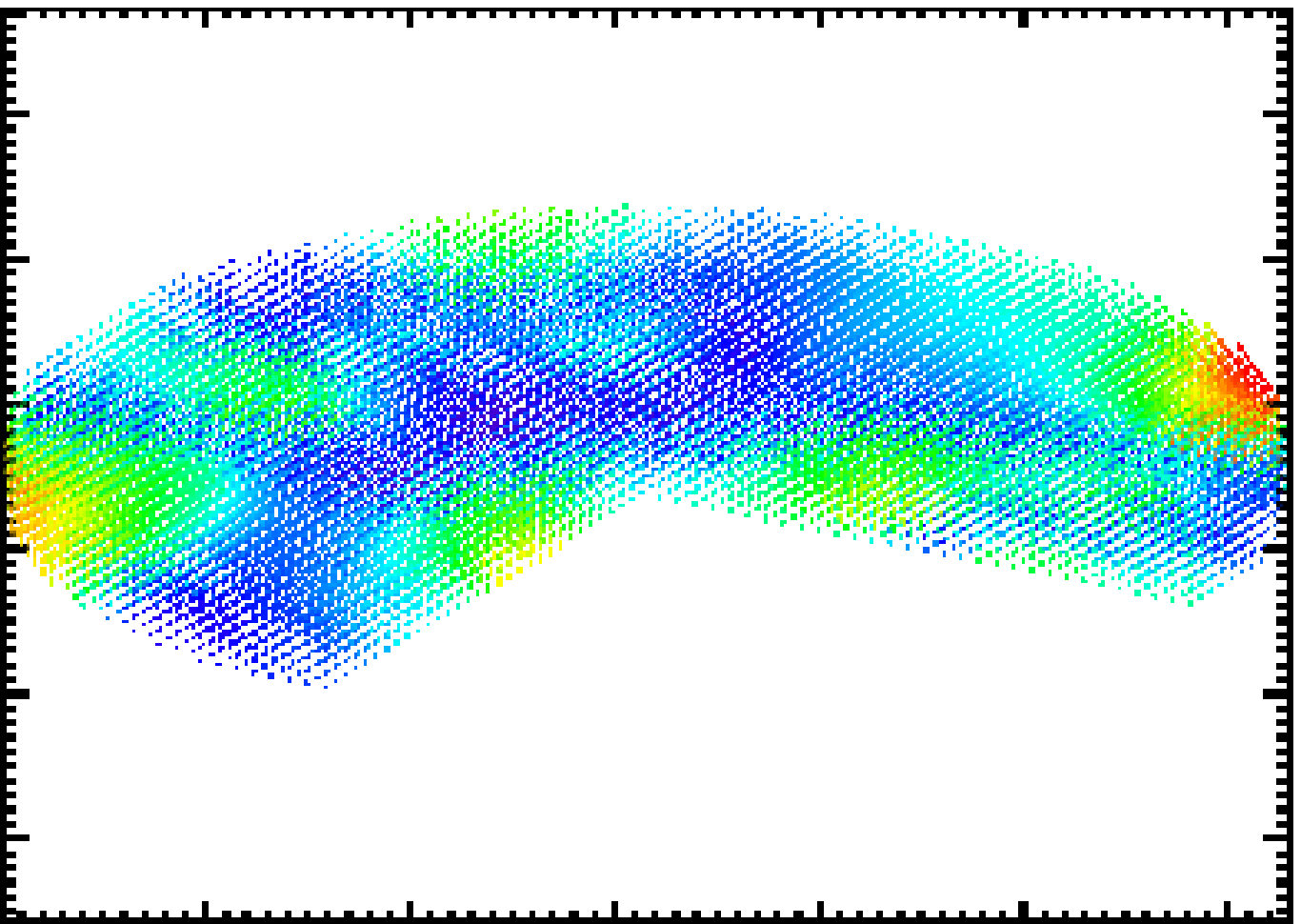}
\end{center}
\caption{\label{fig:label7}  Three-dimensional vector plot of the displacements of the points representing the glass cylindrical optic.  The colors are proportional to the magnitude of the displacement (purple=smallest and red=largest).  The optic is supported from a set of staggered points at the bottom surface (7 points across).  The points are then displaced from their ideal position by a gaussian distribution with standard deviation of 100 nm.}
\end{figure*}

To validation the numerical approach, we first compare with two analytic solutions for a circular plate.  We simulate a cylindrical plate with a radius of 500 $mm$ and a thickness of 100 $mm$.  We assume the plate is made of fused silica glass with a heat capacity of $800 J K^{-1}$, density of 2700 $kg m^{-3}$, a poisson ratio of $0.19$, a Young's modulus of $7 10^{10} Pa$, and a thermal expansion coefficient of $9 10^{-6}$.  The plate is mounted from the side by 10 mount points.

We first test the thermal expansion by cooling the plate 10 degrees below the temperature it was constructed.  Figure~\ref{fig:label1} shows the displacement of the optics from the nominal lattice points.  The thermal expansion of a circular plate follows

$$ \delta h = \alpha h_0$$
$$ \delta r = \alpha r_0$$

\noindent
where $\delta h$ and $\delta r$ are the vertical and radial displacements, $h_0$ and $r_0$ are the original vertical and radial dimensions, and $\alpha$ is the coefficient of thermal expansion.  In Figures~\ref{fig:label2} and~\ref{fig:label3} we demonstrate the consistency of the expansion with the analytic prediction.  We vary the numerical resolution for N $=$ (16, 24, 32, 48, 64, 96, 128).  For all numerical resolutions, the final displacements are within 10 $nm$ from the prediction.

To test the elasticity calculation, we calculate the gravitational sag for the same plate oriented parallel to the horizon.  Figure~\ref{fig:label4} shows the displacements due to the gravitation field.  Poisson found the solution the bending of circular plates with supported edges as

$$ z (r) = -\frac{3mg (1- \nu^2)}{16 \pi r_0^2 E h_0^3} ( r_0^2 - r^2)^2$$

\noindent
where $r$ is the radius, $m$ is the mass, $g$ is the gravitational acceleration, $E$ is Young's modulus, $r_0$ is the radial size, and $h_0$ is the vertical size.  Figure~\ref{fig:label5} shows the agreement with this prediction at various numerical resolutions for N $=$ (16, 24, 32, 48, 64, 96, 128).  For $N$ greater than 64, the agreement is within 25 $nm$ and it converges to the result asymptotically.  We also demonstrate the effect of a thermal gradient on the optic in Figure~\ref{fig:label6} and the effect of the displacements of actuators if it is mounted from the bottom instead of the sides.  We therefore have a robust code for estimating the distortion of the 3-D structure of optics.

\section{Control Systems}

For optics (most likely a mirror) that are actively controlled, a large control loop needs to be used to repeat the calculation above.  The presence of an actuator and its possible non-standard position will modify the overall shape of the optic.  In order to describe this control loop, we:  1) estimate how different the optics shape from its ideal location, 2) add possible observational errors to this surface, 3) iterate the actuator positions to correct the surface, 4) add actuator positioning errors, and 5) adjust the focus position of the optic.   The iterative surface estimation and actuator correction loop (step 2-4) is equivalent to a closed loop active optics system when that part of the simulation is used.  Similarly, the overall focus correction (step 5) due to thermal expansion and the fact that the overall gravitational support is effectively cancelled in our setup is equivalent to the most important corrections of an open loop active optics system.  This control loop is described below.

We first calculate the variation of the derivative of the error of the surface of the mirror.  We first find all the node points that will be considered part of the optical surface.  We then calculate the numerical derivative in both the x and y components.  To do this, we first estimate the local surface error for each of the node points.  This is complicated by the fact that the node points move in not only $z$ direction but also $x$ and $y$.  We therefore estimate the surface error, $\Delta z$ as

$$ \Delta z = z^{\prime} - z - \frac{dh}{dx} (x^{\prime} - x) - \frac{dh}{dy} (y^{\prime} - y) $$

\noindent
where the unprimed coordinates are the original positions (prior to thermal expansion/contractions) and the primed coordinates are the current node coordinates.  $\frac{dh}{dx}$ and $\frac{dh}{dy}$ are the numerical derivatives of the asphere surface.   To calculate the derivative of this displaced surface we then find the displaced surface by using a weighted set of 3x3 points.  We ignore points where the derivative will be using points off the optical surface.  We then compute the average radial derivative by projecting both the x and y components.  We subtract off the expected defocus distortion of $r^2/r_0 \alpha \Delta T$ where $r_0$ is the radius of curvature, $alpha$ is the thermal expansion coefficient, and $\Delta T$ is the change in soak temperature.  This then does not attempt to correct defocus of an ideal parabola, since this can much more easily be accommodated by a focus change.  The focus correction can be empirically determined for optics that deviate from a parabola significantly.  We then compute the average variation of the error surface derivative after this correction and minimize this in the following loop.

We also consider that with most modern telescopes there will be an error in estimating how distorted the surface (wavefront sensing error), depending on both the technology as well as the control system logic.  For that reason, we add an error in the position to each point on the surface in the calculation of $\Delta z$ above.  For ground-based systems, that estimate the distortion this will be dominated by a combination atmospheric and dome/mirror seeing noise.  In that case, we can add a systematic distortion across the surface and have its shape follow the zernike expansion of a Kolmogorov atmosphere ( \citealt{winker1991}, \citealt{boreman1996}).  The overall wavefront sensing error is then scaled to the current seeing in the simulation.  More complicated representations of the wavefront error can also be represented by an additional error on the surface having some spatial structure by simulating the exact method of wavefront sensing.

Given the variation of the surface derivative, we now want to minimize it by applying motions to the actuators.  We experimented with several methods for doing this.  Since there may be hundreds of actuators and a very non-linear mapping from each actuator offset to the reduction of the variation of the surface derivative, we used a Markov Chain Monte Carlo (MCMC) approach (see e.g. \citealt{gilks1996}).  This is well-suited to minimizing large numbers of parameters, particularly where we do not need perfect optimization.  We also experimented with how to do the parameterization of the MCMC.  One method is simply to have each actuator's position a free parameter.  However, it works better to group the position by common radial functions since many of the surface corrections will be large-scale patterns and not small deformations just related to an actuators region of influence.  Therefore, we use a large number of zernike polynomials and then perturb each actuator according to the zernike expansion.  The free parameters are then the zernike coefficients and we use 91 terms.

To perform the MCMC loop, we consider new coefficients and use the standard Metropolis-Hastings acceptance criterion (\citealt{metropolis1953}, \citealt{hastings1970}).  The likelihood is constructed by taking $\exp{-\frac{D^2}{\sigma^2}}$ where $D^2$ is the variance of the surface derivative discussed above and $\sigma$ controls the numerical accuracy.  We found that choosing $\sigma = 0.05" \frac{N}{64}$ achieves the reasonable results.  We use the first 100 iterations as a burn-in phase and scale $D^2$ with a coefficient that goes from 0 to 1.  The initial proposal step is chosen to be from a gaussian distribution centered at 0 with width $\frac{0.2}{o} \mu m$ where $o$ is the order of the zernike polynomial.   After 100 iterations, we use an adaptive step proposal as in \citealt{peterson2007} where we adjust the proposal width to be equal to the standard deviation of the last 100 iterations of the accepted parameter in the Markov Chain.   During each iteration we let the physics simulation respond to the possible actuator changes and settle into a new equilibrium.
To be conservative and avoid any hysteresis-like effect because of the previous iteration, we repeat the calculation of the likelihood for both the new step and the current step when assessing the Metropolis-Hastings criterion.  Thus, we make the Markov Chain half as efficient.  We also make sure the final rms surface height change is a small fraction ($0.1\%$ typically) of the difference in the surface height between the Markov steps.  We let the physics simulation continue if this is not the case.  At the end of the process, we then have an estimate of the final actuator positions which minimizes the variance of the surface error derivative.  Finally, after minimization we add a possible actuator positioning error.  Every actuator is pushed by an amount equal to the expected positional tolerance.

\section{Interface to Ray-Traced Surface}

It is not simple to match the output of the surface deformation calculation with that of the photon raytrace.  This because the surface deformation calculation is a 3-D lattice of three-dimensional vector displacements of modest resolution (~$64$).  The optical raytrace needs a two-dimensional map of a vertical displacement of high resolution (~$1024$).  To match the two, we use the following procedure.

First, we select the subset of points of the 3-d lattice that represent the optical surface.  Then for those points we have a list with the original position ($x_0$, $y_0$, $z_0$) and their final displacements ($\Delta x$, $\Delta y$, $\Delta z$).  For each of these points, we want to combine the information in the lateral displacements ($\Delta x$, $\Delta y$) and the vertical displacement ($\Delta z$) into a single vertical displacement.  So we compute the perturbed radial coordinate $r^{\prime} =\sqrt{(x+\Delta x)^2 + (y + \Delta y)^2}$, and then compute the surface height from the asphere equation,  $h(r^{\prime})$.  Then the net surface displacement compared to the unperturbed surface is $z_0 + \Delta z - h(r^{\prime})$.

We then want to match these points with the much larger number of points that currently represent the optical surface of the given optic.  We use the Fast Library for Approximate Nearest Neighbors  (\citealt{blanco2011}, \citealt{muja2009}) to find the near neighbors of a given optical surface point.  We use the nearest $\sqrt{\frac{1}{2} N}$ points, where $N$ is the subset of lattice points.  Then, we experimented with a few different approaches to fitting surface functions or using inverse distance weighting to use these points.  We found the most robust results with simply performing a linear fit.  Then this process is repeated for the roughly one million points that represent the optical surface.  This results in a perturbed surface at arbitrarily fine resolution, and minimal artifacts.  We tested the convergence of this procedure by varying the number of lattice node points.

\section{Application to the WIYN 3.5m Telescope}

\begin{figure*}[htb]
\begin{center}
\includegraphics[width=0.99\columnwidth]{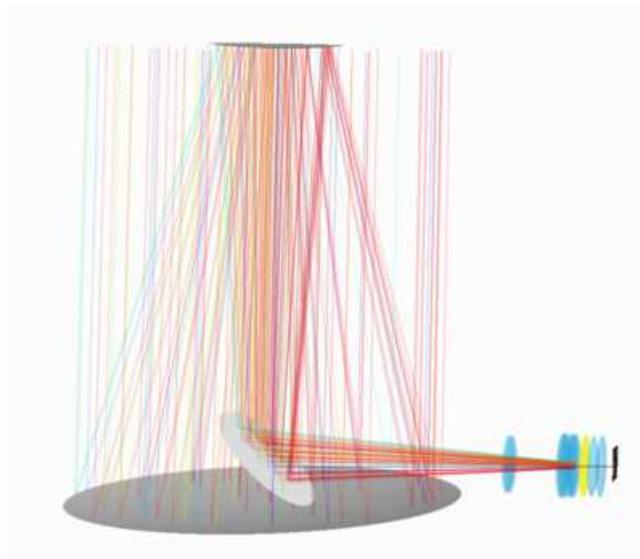}
\end{center}
\caption{\label{fig:label8}  The optical configuration of the WIYN 3.5m telescope as implemented in PhoSim.  The grey indicates the position of the mirrors, the blue indicates the lens surfaces, and the yellow indicates the filter surfaces.  The red lines indicate the paths of the photons.}
\end{figure*}

\begin{figure*}[htb]
\begin{center}
\includegraphics[width=0.99\columnwidth]{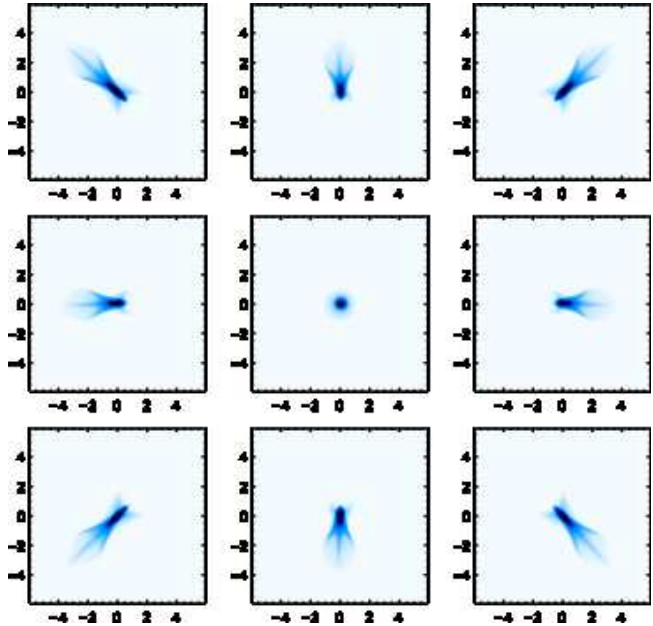}
\end{center}
\caption{\label{fig:label9}  Spot diagrams of the point spread function at the centers of the central 3x3 chips of the ODI focal plane.  Darker blue indicates higher intensity and the scale is in microns.  Here there are no perturbations on the mirror, so the diagram represents the intrinsic aberrations of the WIYN telescope. }
\end{figure*}

\begin{figure*}[htb]
\begin{center}
\includegraphics[width=0.99\columnwidth]{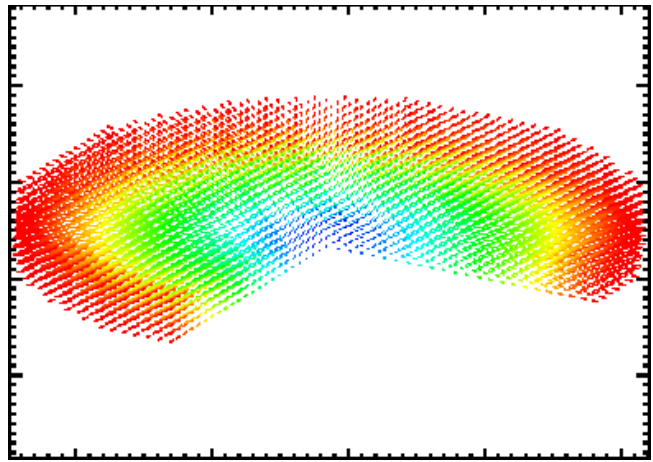}
\end{center}
\caption{\label{fig:label10}  The effect of a 20 degree soak temperature change on the WIYN primary mirror.  The vectors indicate the direction of the displacement, and the color indicates the magnitude (red=larger and purple=smaller).}
\end{figure*}

\begin{figure*}[htb]
\begin{center}
\includegraphics[width=0.99\columnwidth]{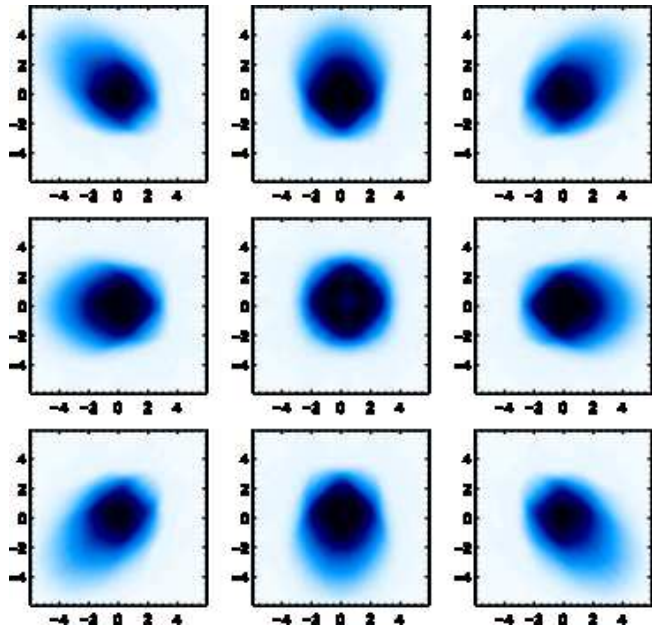}
\end{center}
\caption{\label{fig:label11}  Spot diagrams of the point spread function at the centers of the central 3x3 chips of the ODI focal plane with a change in soak temperature of 20 degrees.  Darker blue indicates higher intensity and the scale is in microns.  The optics have been corrected for the large-scale defocus, but there are some residual focal errors.}
\end{figure*}

\begin{figure*}[htb]
\begin{center}
\includegraphics[width=0.99\columnwidth]{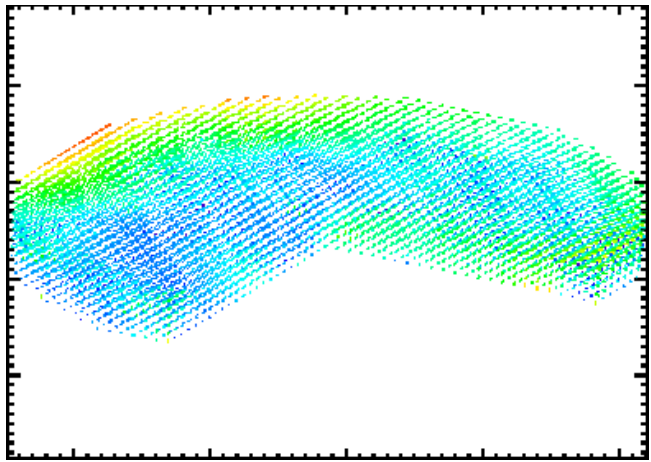}
\end{center}
\caption{\label{fig:label12}  The effect of the gravitational sag when the mirror is tilted at 30 degrees above the horizon and the actuator positioning errors on the WIYN primary mirror.  The vectors indicate the direction of the displacement, and the color indicates the magnitude (red=larger and purple=smaller). }
\end{figure*}

\begin{figure*}[htb]
\begin{center}
\includegraphics[width=0.99\columnwidth]{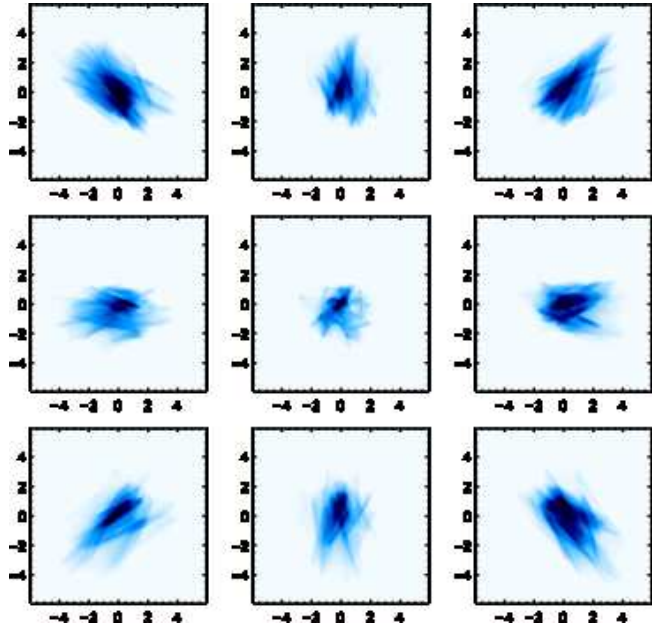}
\end{center}
\caption{\label{fig:label13}  Spot diagrams of the point spread function at the centers of the central 3x3 chips of the ODI focal plane with gravitational sag from 30 degrees above the horizon and actuator positioning errors.  Darker blue indicates higher intensity and the scale is in microns.  The diagrams indicate fine structure features resulting from gravitational sag between the mount points.}
\end{figure*}

\begin{figure*}[htb]
\begin{center}
\includegraphics[width=0.99\columnwidth]{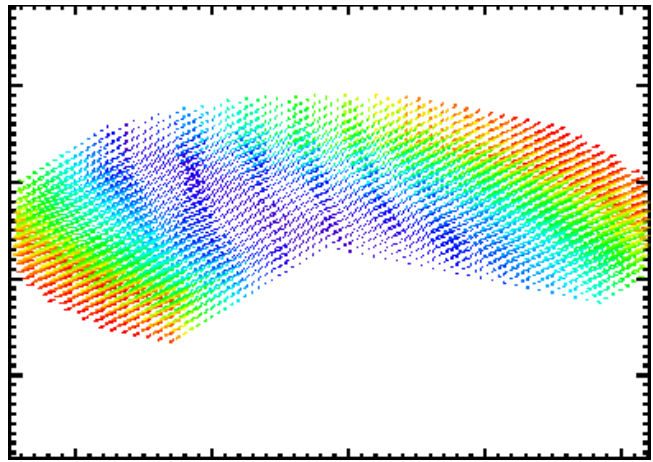}
\end{center}
\caption{\label{fig:label14}  The effect of a thermal temperature gradient on the WIYN primary mirror.  The vectors indicate the direction of the displacement, and the color indicates the magnitude (red=larger and purple=smaller).}
\end{figure*}

\begin{figure*}[htb]
\begin{center}
\includegraphics[width=0.99\columnwidth]{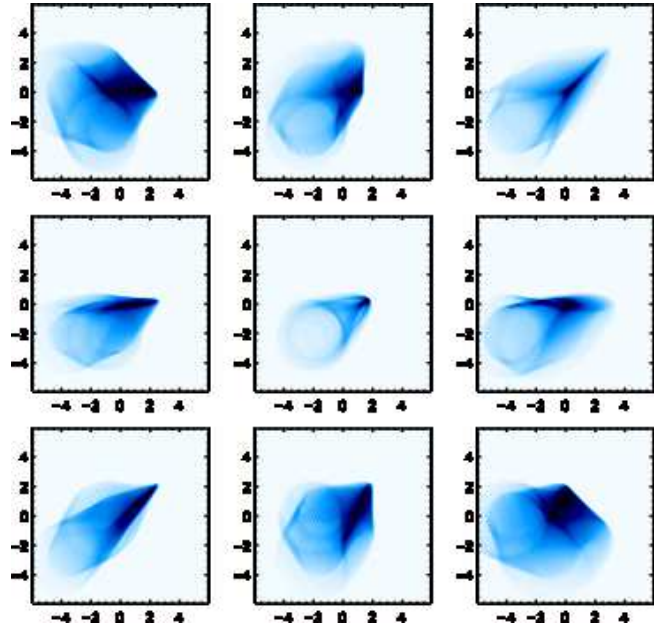}
\end{center}
\caption{\label{fig:label15}  Spot diagrams of the point spread function at the centers of the central 3x3 chips of the ODI focal plane with a thermal gradient across the mirror.  Darker blue indicates higher intensity and the scale is in microns.  The diagram indicates a structured pattern across the field from the thermal distortion across the mirror.}
\end{figure*}

\begin{figure*}[htb]
\begin{center}
\includegraphics[width=0.99\columnwidth]{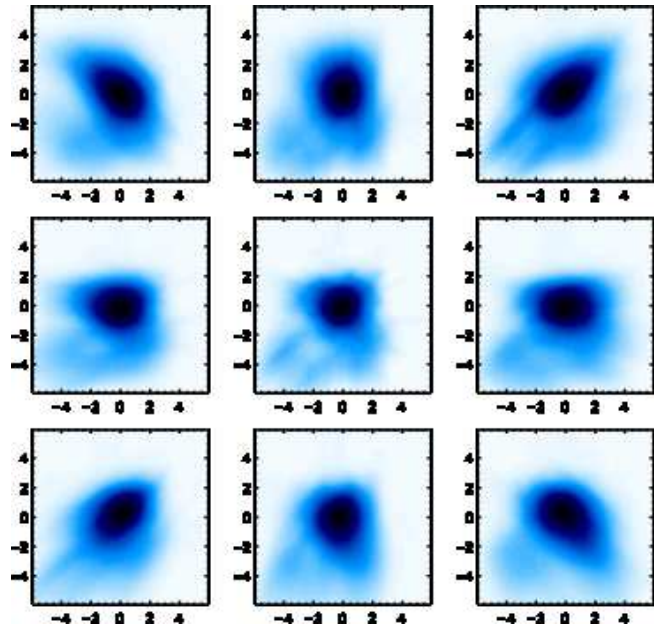}
\end{center}
\caption{\label{fig:label16}  Spot diagrams of the point spread function at the centers of the central 3x3 chips of the ODI focal plane with a soak temperature change of 20 degrees, a thermal gradient across the mirror, and gravitational sag from the mirror tilted 30 degrees above the horizon, and actuator positioning errors.  Darker blue indicates higher intensity and the scale is in microns.  The diagrams are effectively the combination of three previous figures.}
\end{figure*}

\begin{figure*}[htb]
\begin{center}
\includegraphics[width=0.99\columnwidth]{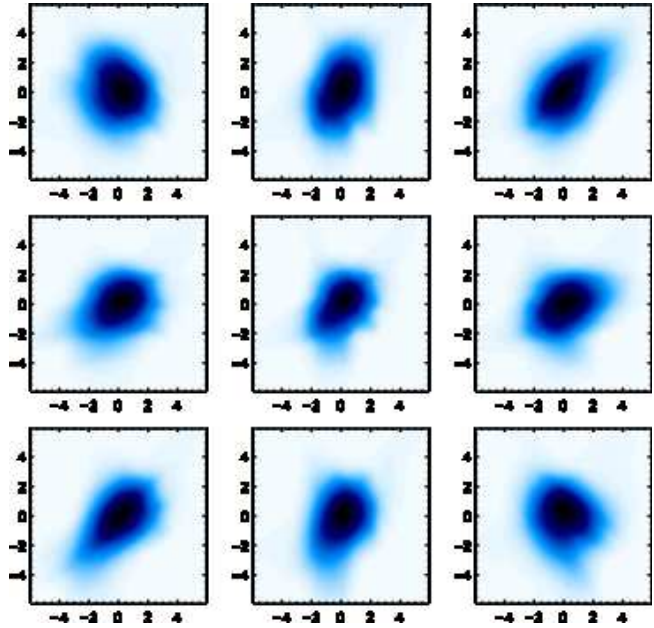}
\end{center}
\caption{\label{fig:label17}  Spot diagrams of the point spread function at the centers of the central 3x3 chips of the ODI focal plane with a soak temperature change of 20 degrees, a thermal gradient across the mirror, and gravitational sag from the mirror tilted 30 degrees above the horizon, and actuator positioning errors.  These simulations use the control loop to modify the surface to minimize the derivative variation.  Darker blue indicates higher intensity and the scale is in microns.}
\end{figure*}

\begin{figure*}[htb]
\begin{center}
\includegraphics[width=0.99\columnwidth]{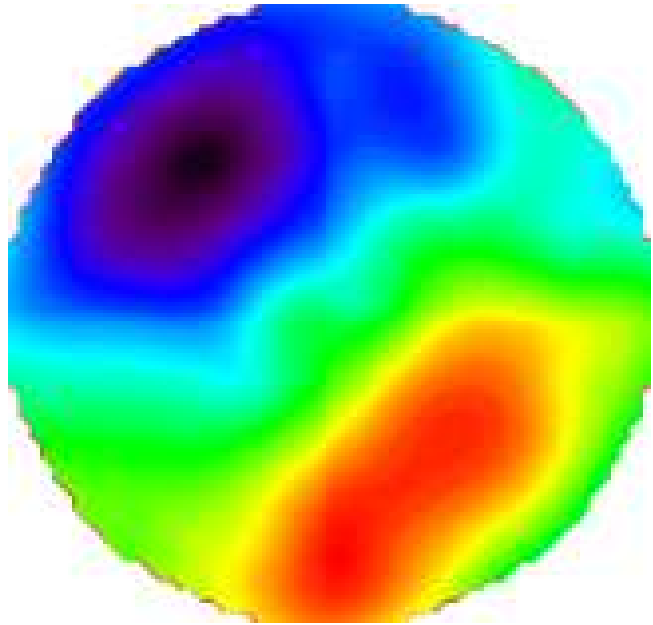}
\end{center}
\caption{\label{fig:label18} The surface map of the primary mirror from the same simulation in Figure~\ref{fig:label16}.  This simulation uses a soak temperature change of 20 degrees, a thermal gradient across the mirror, and gravitational sag from the mirror tilted at 30 degrees above the horizon and positioning errors.   We have removed the main defocus distortion as well as the second order radial pattern.  The scale is $\pm$ 100 nm.  }
\end{figure*}

\begin{figure*}[htb]
\begin{center}
\includegraphics[width=0.99\columnwidth]{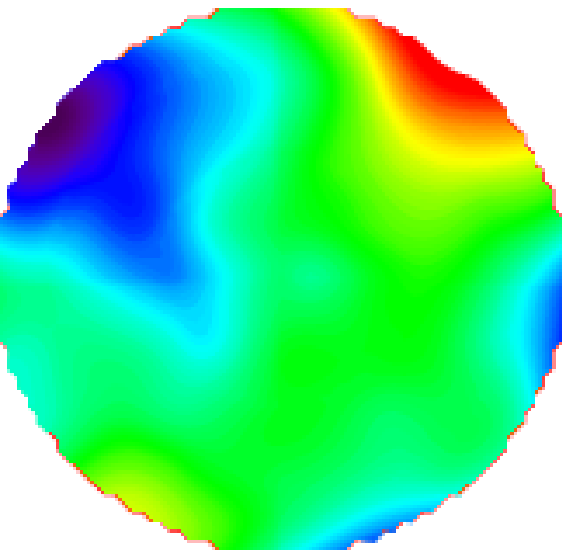}
\end{center}
\caption{\label{fig:label19}  The surface map of the primary mirror from the same simulation in Figure~\ref{fig:label17}.  This simulation uses a soak temperature change of 20 degrees, a thermal gradient across the mirror, and gravitational sag from the mirror tilted at 30 degrees above the horizon, positioning errors, and with the control system correction.   We have removed the main defocus distortion as well as the second order radial pattern.  The scale is $\pm$ 100 nm.  Compare this to the Figure~\ref{fig:label18} to see the effect of the control system.}
\end{figure*}

The WIYN 3.5m telescope (\citealt{johns1994}, \citealt{johns1990}) is a well-studied system that contains a large primary honeycomb-structured borosilicate mirror (\citealt{angel1984}) that is actively controlled (\citealt{roddier1995}).  We implemented the WIYN telescope in PhoSim with the One Degree Imager (ODI) focal plane (\citealt{jacoby2002})).  The optical design was adapted from http://www.wiyn.org/About/Wiyn3.5mTelescope.zmx.  For the ODI focal plane we adapted the specifications of http://www.wiyn.org/ODI/index.html.  For this work, we consider the thermal and  mechanical perturbations to the primary mirror.  For the borosilicate, we use a density of 2230 $kg~m^{-3}$, a heat capacity of $800 J C^{-1}$, a thermal conduction coefficient of 1.2 $W m^{-2} C^{-1}$, a thermal expansion coefficient of $3.3 10^{-6}$, and a Young's Modulus of $6.3 10^{10} Pa$.  We use an edge thickness of 463.9 $mm$ and a fill factor of 23$\%$ to reproduce the weight of 1964 $kg$ according to the specifications at http://www.wiyn.org/About/wiynspecs.html.  The actuators are placed in approximately the same pattern with a region of influence of 10 $cm$ and a rms positioning error of 25 $nm$.  We apply a cooling performance rate of $160 W C^{-1}$ by using the specifications of \citealt{goble1991} and assuming a coupling efficiency of 20$\%$.
 All of our simulations of WIYN are equivalent to short exposures (less than a minute) where there is no significant change of the environment during the exposure that would affect the temperature, temperature gradients, or gravitational forces.  Longer exposures can be generated by co-adding a time sequence of the environmental conditions.

Given the implementation, we then simply run PhoSim with the thermal-elastic calculation coupled to the full optical raytrace.  For the following examples, we use raytracing in the pure geometric optics limit.  This neglects a contribution diffractive effects, but makes comparisons for straight-forward for WIYN (see e.g. \citealt{harmer2002}).  The actual PSF from the various optics configuration would be modestly larger than the following examples.  With the nominal conditions of the mirror pointed at zenith, the mirror soak temperature being at the same temperature it was figured, no thermal time gradient, and the control system off, we expect no displacements of the mirror.  We then compute the point spread function of the optics alone by turning off any effect of the atmosphere and the detector.  Figure~\ref{fig:label8} shows the implementation of the WIYN optical design and the typical ray trajectories.  We compute spot diagram by simulating 9 sources on a 3 x 3 grid separated by 0.13 degrees with a 14th magnitude.  The pixels are reduced by a factor of 100 in order to see the PSF structure which mostly occurs within about 1 pixel.  The corresponding spot diagrams are shown in Figure~\ref{fig:label9}.  The intrinsic optical aberrations of the optical design are about 0.011 arcseconds.

We then change various aspects of the environment to predict the change the primary mirror shape and the corresponding effect on the spot diagrams.  We first drop the temperature 20 degrees below the temperature the primary mirror was figured.  This results in the thermal expansion of the mirror as shown in Figure~\ref{fig:label10}.  The main effect on the optical system would be to change the focus, which is already included in our formalism above.  However, there are smaller second order effects of this which are not fully corrected and the spot diagrams indicate a typical size of about 0.037 arcseconds as shown in Figure~\ref{fig:label11}.

The effect of gravitational sag is then simulated by tilting the mirror so it is 30 degrees above the horizon.  The actuators correctly keep the mirror in place, but have a positioning error.  The mirror then sags between the actuators according to the gravitational vector and the elastic response.  The displacements from this configuration are shown in Figure~\ref{fig:label12}.  The corresponding spot diagrams are shown in Figure~\ref{fig:label13}.  The highly structured pattern is then due to the higher frequency distortions on the mirror surface.  The average PSF size is 0.023 arcseconds.

We then explore the effect of a thermal gradient.  We set the temperature change to a large value of 2 degrees per hour.  This then predicts a magnitude of the temperature gradient according to our methodology in \S3.  The displacement of the primary mirror is shown in Figure~\ref{fig:label14}.  The corresponding spot diagram pattern is shown in Figure~\ref{fig:label15}.  The effect of the temperature gradient is to cause a significant elliptical shape in a common direction with some field dependence.  The average PSF size is 0.028 arcseconds.

Finally, we include all three effects of the environment in Figure~\ref{fig:label16}.  There is some defocusing from the soak temperature, structured PSF blurring from the gravitational sag and positioning errors, and some common ellipticity distortion from the thermal gradient.  The average PSF size is 0.037 arcseconds.  We then repeat this calculation but allow the whole control loop to try to correct the surface and minimize the variation of the derivative.  We include a rms wavefront error of 100 nm using the methodology in \S5.  After the loop is completed, the adjusted surface is used to predict the spot diagrams in Figure~\ref{fig:label17}.  As expected the PSF size is decreased to 0.033 arcseconds.  Figures~\ref{fig:label18} and~\ref{fig:label19} show the change in surface shape before and after the control system calculation.  We subtracted the overall distortion of the change in soak temperature in order to see the differences by using the ideal parabola prediction and removing a second order radial polynomial residual function.

All of these spot diagrams are relatively small compared to the typical seeing size at Kitt Peak, the site of WIYN.  By including the misalignments of all the optics, as well as surface pattern of the other optics the PSF size would increase.  The ellipticity of the PSF due to these effects, however, could still be significant even if the PSF size is small.  If a contribution to the PSF from say optical perturbations has a size of $\sigma$, but the total final PSF size is $\sigma_T$, then the ellipticity will get diluted by a factor of ${\sigma^2} / {\sigma_T^2}$.  This is not negligible since the ellipticities of all the PSF patterns presented are between 25$\%$ and 75$\%$.  For 0.5 arcsecond seeing, then this would get diluted by two orders of magnitude, but then would still be comparable to the ellipticity due to the atmosphere from incomplete turbulence averaging.  For lower f-number telescopes, the effect may be larger and will be studied in future work.

\section{Conclusion and Future Work}

The new capabilities in this work should provide a robust methodology to accurately predict realistic PSF patterns and their variation by coupling the deformation of optical surfaces to thermal changes and mechanical stresses.  The code described above is implemented in PhoSim v5.0.2 and available at the public site (https://bitbucket.org/phosim/phosim\_release) \citealt{peterson2018}.  The aberrations predicted are responsible for a significant fraction of the PSF ellipticity and its corresponding variation across the field.  In addition, further insight into the thermal patterns on mirrors could improve the realism of these predictions.  In future work, simulating a suite of telescopes and comparing with actual observations will enable a large number of future studies.

\acknowledgments

JRP acknowledges support from Purdue University.  We thank Garrett Jernigan, George Angeli, Chuck Claver, Bo Xin, and Eugene Magnier for helpful conversations and discussions related to this work.  We thank the anonymous referee for helpful comments.

\end{document}